\documentclass[conference,a4paper]{IEEEtran}
\IEEEoverridecommandlockouts

\usepackage{cite}
\usepackage{amsmath,amssymb}
\usepackage{graphicx}
\graphicspath{{figures/}}
\usepackage{booktabs}
\usepackage{url}
\usepackage{xcolor}
\usepackage{tikz}
\usetikzlibrary{arrows.meta, positioning, calc, fit, backgrounds}

\newcommand{\PLACEHOLDER}[2]{%
  \fbox{\parbox[c][#2][c]{\dimexpr\linewidth-2\fboxsep-2\fboxrule\relax}{%
    \centering\footnotesize\textcolor{red}{PLACEHOLDER: #1}}}}

\begin{document}

\title{A Ray-Traced Digital Twin for 5G NR Positioning with OpenAirInterface}

\author{
\IEEEauthorblockN{
  Nima Chaharbaghi\IEEEauthorrefmark{1},
  Florian Kaltenberger\IEEEauthorrefmark{2},
  Bartosz Podrygajlo\IEEEauthorrefmark{3},
  Rakesh Mundlamuri\IEEEauthorrefmark{3},
  Roberto Verdone\IEEEauthorrefmark{1}
}
\IEEEauthorblockA{\footnotesize
  \IEEEauthorrefmark{1}University of Bologna, Italy \quad
  \IEEEauthorrefmark{2}EURECOM, France \quad
  \IEEEauthorrefmark{3}OAI Software Alliance, France\\[2pt]
  nima.chaharbaghi@studio.unibo.it \quad
  florian.kaltenberger@eurecom.fr \quad
  roberto.verdone@unibo.it\\
  bartosz.podrygajlo@openairinterface.org \quad
  rakesh.mundlamuri@openairinterface.org
}
}
\maketitle

\begin{abstract}
Evaluating 5G positioning on a physical testbed ties every measurement to
one propagation environment and one antenna layout, so any question about
an alternative deployment requires a new field campaign. We present a
ray-traced digital twin of the GEO-5G testbed at EURECOM that drives an
unmodified OpenAirInterface (OAI) 5G stack with per-antenna channels
computed by Sionna~RT over a 3D model of the site. The distributed gNB
antennas are modelled as independent radio devices rather than as
elements of a compact array, which preserves the per-antenna absolute
delays on which time-difference-of-arrival (TDoA) positioning depends.
With them preserved, the twin drives OAI's SRS-based UL-TDoA pipeline to
metre-level accuracy consistent with that reported for the physical
deployment. We then use the twin to answer a question the physical
testbed cannot. Widening the carrier truncates the tail of the horizontal
error distribution without improving a typical position, and leaves the
vertical error untouched, because the latter is limited by the array
geometry rather than by the measurement. That limit is predictable from
the antenna coordinates alone, and a height-diverse layout reduces the
RMS height error by a factor of four.
\end{abstract}


\section{Introduction}

Positioning is one of the capabilities that distinguishes 5G NR from its
predecessors, and open-source implementations of the standardised methods
have now reached the point where they can be evaluated on real
deployments~\cite{malik2025concept,ahadi2025experimental,mundlamuri2025wons}.
That evaluation is costly. It demands spectrum licensing, certified
hardware and a prepared site, and a campaign consumes weeks before a
single measurement is taken. What it returns is narrow: every measurement
is bound to one propagation environment, one antenna layout and one
mobility scenario, and the ground truth carries a survey uncertainty of
its own. Questions of the form ``what would this deployment achieve if
the antennas were placed differently'' cannot be answered without
rebuilding the deployment.

A digital twin removes both constraints. If the channel is computed from
a 3D model rather than measured over the air the environment becomes an
input, and if the user equipment (UE) position is scripted rather than
surveyed the ground truth becomes exact. Ray-traced twins coupled to a
live protocol stack have been demonstrated for 5G
mobility~\cite{iye2025owdt}, and OAI ships a ray-tracing channel emulator
connecting Sionna~RT to its virtual radio
simulator~\cite{raytracingchannelemulator}. Neither addresses
positioning, and both target a single-antenna link, precisely the
configuration in which time-difference-of-arrival (TDoA) has nothing to
measure.

This paper closes that gap and then exploits it. We build a ray-traced
digital twin of the GEO-5G testbed at EURECOM, a distributed eight-antenna
5G deployment on which OAI's uplink TDoA (UL-TDoA) pipeline has already
been evaluated in hardware~\cite{ahadi2025experimental}, and use it to
drive the same unmodified pipeline. Our contributions are:

\begin{enumerate}
  \item \textbf{A distributed-antenna ray-traced channel emulator},
    extending the OAI ray-tracing channel
    emulator~\cite{raytracingchannelemulator} from one co-located antenna
    to eight spatially separated elements, each an independent radio
    device so that its absolute propagation delay survives into the
    emulated channel. Neither softmodem is modified.

  \item \textbf{A verification against the positioning pipeline the twin
    must support}: over sixteen positions in two scenes and at two
    bandwidths, the emulated channel carries enough delay information for
    the unmodified OAI Location Management Function (LMF) to recover the
    geometry it was generated from, to a median horizontal error of
    0.71--1.20\,m~\cite{ahadi2025experimental}.

  \item \textbf{A deployment finding the physical testbed cannot
    produce}: vertical accuracy is insensitive to a $2.5\times$ increase
    in channel bandwidth because it is limited by the antenna geometry, a
    limit that follows from the coordinates alone, and a height-diverse
    layout reduces the RMS height error by a factor of four while leaving
    the horizontal geometry untouched.
\end{enumerate}

\section{Related Work}

Reference architectures for network digital twins treat GPU-accelerated
ray tracing as the propagation engine of choice~\cite{nokia2025dt}, and a
benchmark of ray-based solvers identifies propagation computation as the
principal obstacle to real-time operation~\cite{zhu2024}. Neither couples
ray tracing to a standards-compliant radio access network.

Closer to this work, an open wireless digital twin couples ray tracing to
OAI to emulate 5G mobility end to end~\cite{iye2025owdt}, and OAI's own
ray-tracing channel emulator connects Sionna~RT to \texttt{vrtsim}
through a scenario-control API~\cite{raytracingchannelemulator}. Both
establish that an unmodified OAI stack can be driven by ray-traced
channels, and both are limited to a single-antenna link, which produces
no arrival-time differences and so cannot exercise a TDoA pipeline.

On the positioning side, a fully open-source 3GPP-compliant UL-TDoA
implementation spanning gNB and LMF has been described~\cite{malik2025concept}
and evaluated on OAI testbeds including GEO-5G, reaching 1--2\,m accuracy
in 90\% of cases and identifying synchronisation and multipath as the
binding constraints~\cite{ahadi2025experimental}. Complementary SDR
measurements characterise TDoA estimation on a smaller
testbed~\cite{palama2024sdr}, round-trip-time estimation has been added to
the same stack~\cite{mundlamuri2024rtt}, and the tooling and methodology
for NR positioning experiments with OAI have been surveyed
in~\cite{mundlamuri2025wons}. All of this work evaluates against deployed
hardware.

The literature therefore covers ray-traced twins of the propagation
environment, OAI integration with a single-antenna ray-traced channel, and
hardware evaluation of OAI's TDoA pipeline, but not their combination.
We supply it.

\section{Digital Twin Architecture}
\label{sec:arch}

\subsection{Reference deployment}

The GEO-5G testbed is EURECOM's outdoor 5G deployment~\cite{ahadi2025experimental}:
an OAI-based virtualised RAN with two commercial O-RAN radio units, one on
the south terrace and one on the north rooftop, each carrying four
spatially separated directional elements. Its eight gNB antenna elements
are therefore not a panel. They form two collinear sub-arrays of four,
sub-array~A spanning 36.0\,m along the south terrace and sub-array~B
spanning 15.0\,m along the north rooftop, with horizontal axes mutually
orthogonal to within a fraction of a degree and horizontal centroid
separation 42.19\,m. We
take the element coordinates from the surveyed placement published with
the EURECOM 5G SRS dataset~\cite{ahadi2025dataset}. The array surrounds a
courtyard rather than looking across it. The carrier is 3.6192\,GHz
(band n78), so the wavelength is 8.28\,cm and the element separations run
from roughly $60\lambda$ to $217\lambda$.

\begin{figure}[t]
  \centering
  \IfFileExists{figures/eurecom_map.png}{%
    \includegraphics[width=\columnwidth]{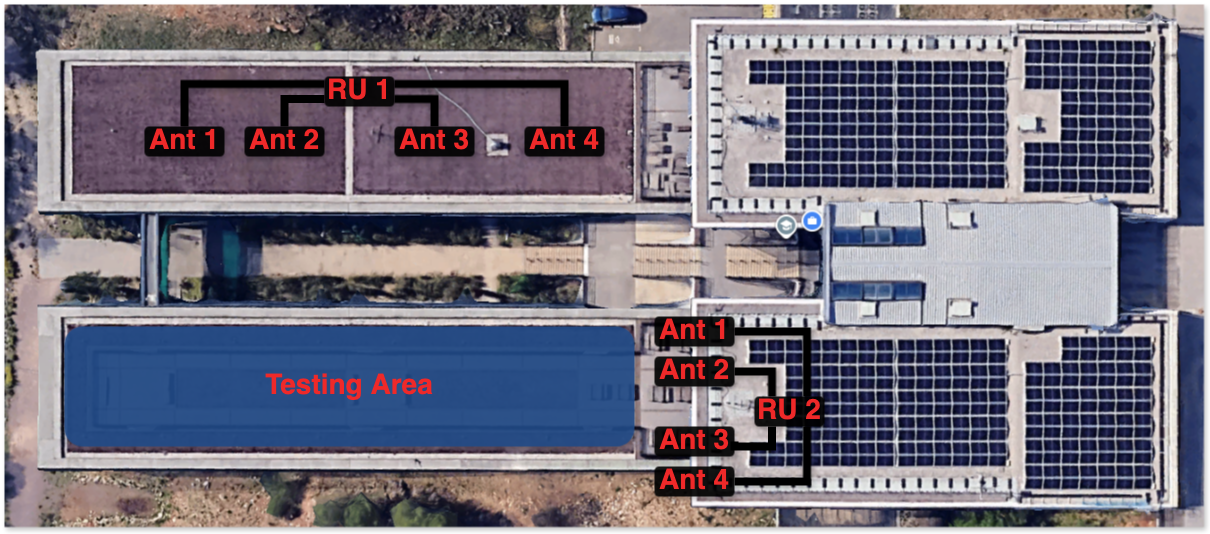}}{%
    \PLACEHOLDER{Fig.~1: sky view of the EURECOM campus with the eight
      antenna elements marked. Export the site map from the dataset
      repository to \texttt{figures/} as \texttt{eurecom\_map.png}.}{42mm}}
  \caption{The eight gNB elements on the EURECOM campus. Sub-array~A runs
    along the south terrace with 9.0, 18.0 and 9.0\,m spacings;
    sub-array~B runs along the north rooftop with 5.0\,m spacings, on an
    axis orthogonal to~A. The two sub-arrays enclose a
    courtyard rather than looking across it, so several test positions
    reach part of the array only by reflection. Element coordinates and
    site map from the EURECOM 5G SRS dataset~\cite{ahadi2025dataset}.}
  \label{fig:site}
\end{figure}

Fig.~\ref{fig:site} shows the placement. That last quantity, the element separation in wavelengths, is the design constraint for everything that follows.

\subsection{Scene and propagation}

The campus is reconstructed from OpenStreetMap
footprints~\cite{osm_data}, refined manually, assigned
frequency-dependent ITU radio materials, and exported to the Mitsuba
format that Sionna~RT
consumes~\cite{hoydis2023sionnart,sionna_rt_techreport}; the scene and
its material assignments are maintained
in~\cite{qin_eurecom_scene}. Paths are found
by Sionna~RT's two-stage solver: shooting-and-bouncing generates candidate
paths, and an image-method stage refines each into an exact specular path.
The solver configuration is held fixed across all experiments:
\texttt{max\_depth}~10, line-of-sight, specular reflection and refraction
enabled, diffuse reflection disabled, SBR seed 41.

\begin{figure}[t]
  \centering
  \IfFileExists{figures/scene_eurecom.png}{%
    \includegraphics[width=\columnwidth]{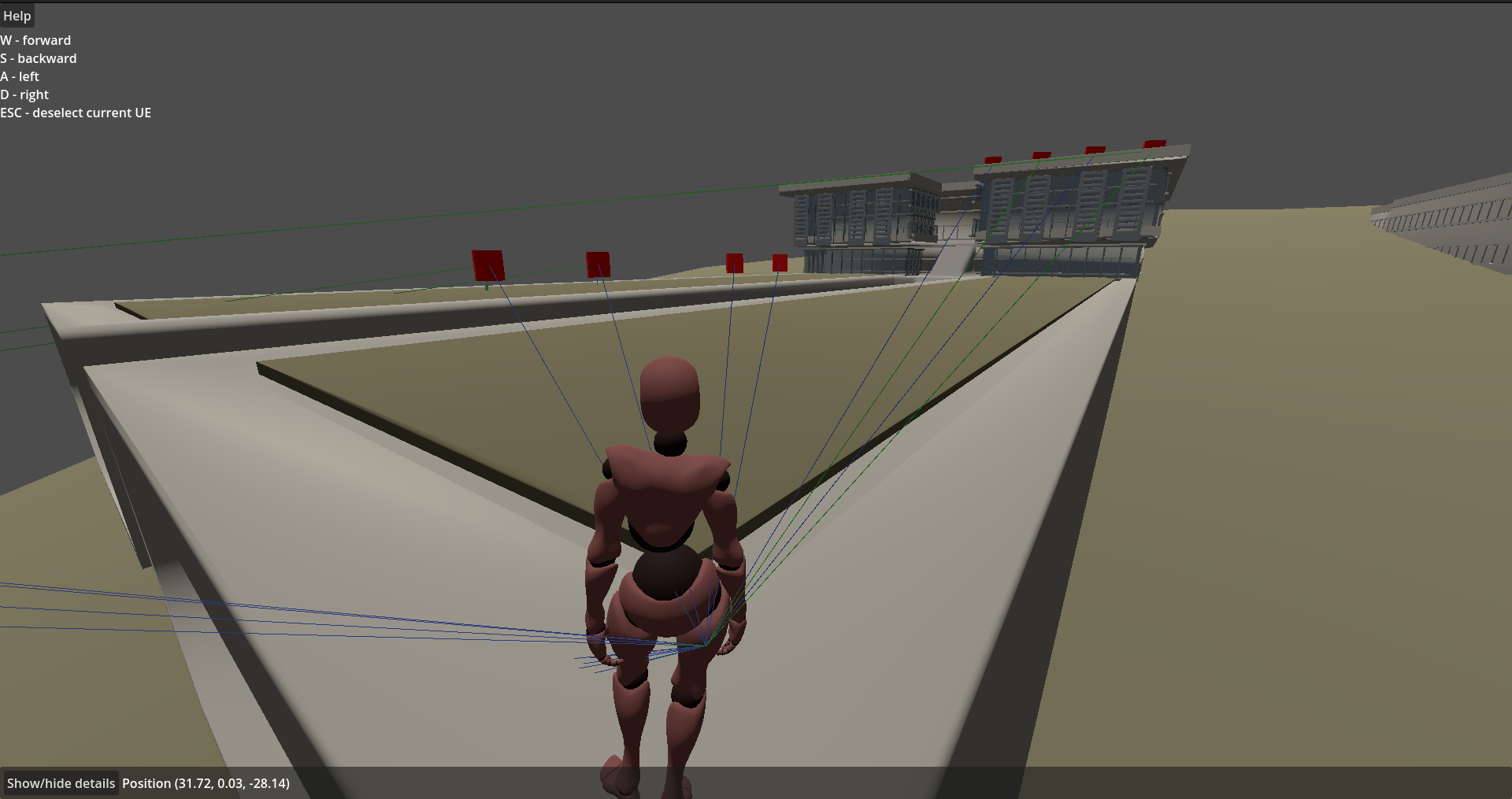}}{%
    \PLACEHOLDER{(a) EURECOM scene}{40mm}}\\[1mm]
  {\footnotesize (a) EURECOM}\\[2mm]
  \IfFileExists{figures/scene_munich.png}{%
    \includegraphics[width=\columnwidth]{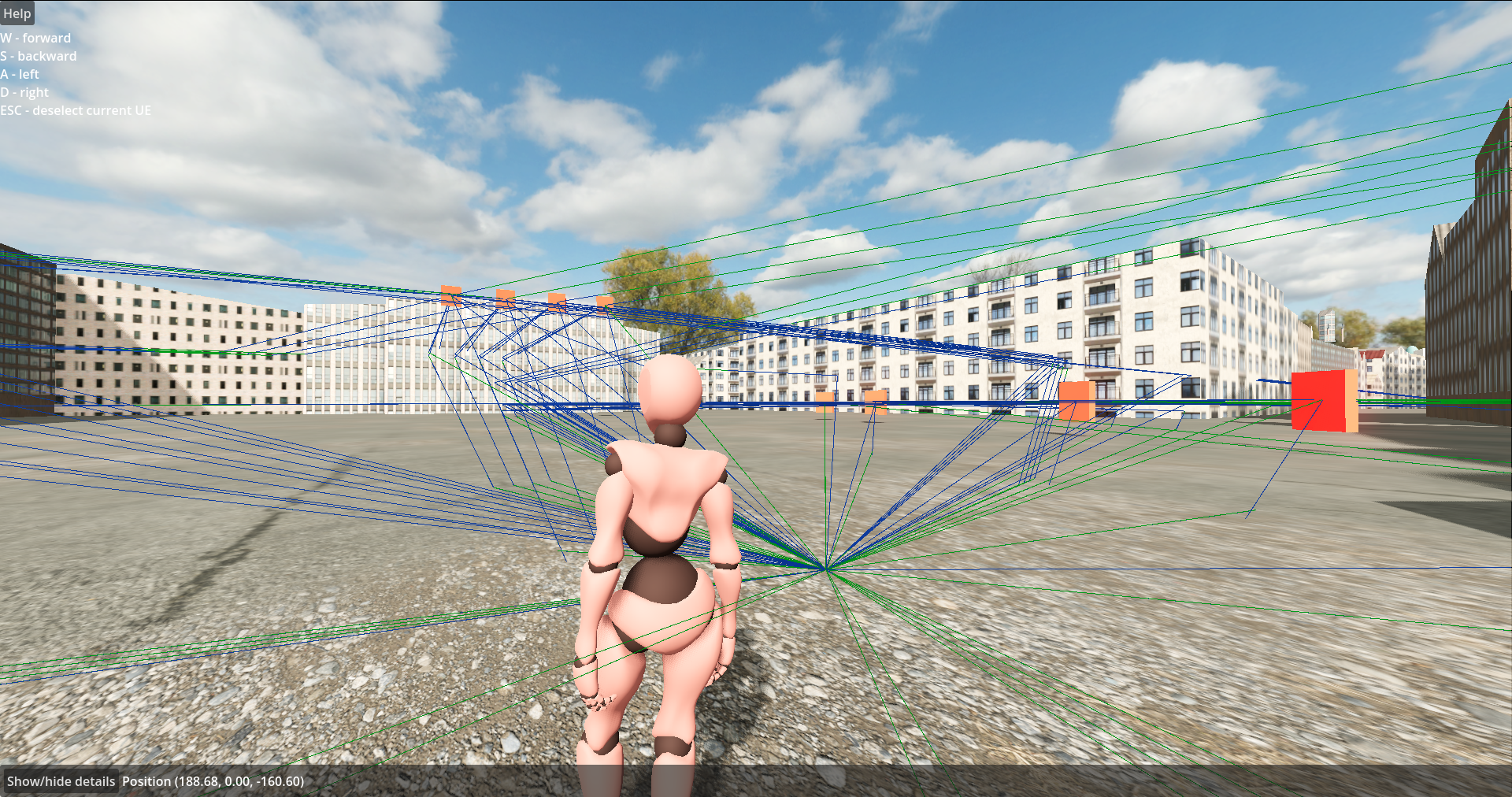}}{%
    \PLACEHOLDER{(b) Munich scene}{40mm}}\\[1mm]
  {\footnotesize (b) Munich}
  \caption{The two ray-tracing scenes, rendered from the same geometry the
    solver uses. (a) The EURECOM campus, reconstructed from OpenStreetMap
    footprints and refined manually; the array encloses the courtyard
    visible at the centre. (b) The Munich scene distributed with
    Sionna~RT, a dense-urban model unrelated to the deployment, in which
    the same array is instantiated under a distance-preserving transform.}
  \label{fig:scenes}
\end{figure}

A depth of ten was chosen because several antenna positions see the UE
only through multiply reflected paths in the partially enclosed
courtyard; we ran no sensitivity study over depth, so it should be read
as a conservative choice rather than a measured requirement. Diffuse
scattering is disabled and the SBR seed fixed so that repeated runs
produce bit-identical taps, any run-to-run variation reaching OAI as
spurious channel variation. The cost is a sparser impulse response than
a measured one (Section~\ref{sec:limits}).
Fig.~\ref{fig:scenes} shows both scenes.

\subsection{Distributed-antenna channel model}
\label{sec:distributed}

Sionna~RT models an array as elements attached to a single radio device
at one position; with \texttt{synthetic\_array} enabled, per-element
responses follow from phase shifts derived under a plane-wave
assumption~\cite{sionna_rt_techreport}. That is correct for a panel a few
wavelengths across and wrong here. Over separations of $60$--$217\lambda$
at ranges of 24--57\,m each element has its own visibility of the UE, its
own set of reflectors and, decisively, its own absolute delay; collapsing
them onto a common phase centre discards exactly the quantity a TDoA
estimator consumes.

We therefore instantiate each element as an independent Sionna radio
device carrying a $1{\times}1$ array: eight \texttt{Transmitter}
objects in the downlink scene and eight \texttt{Receiver} objects in the
uplink scene, with the single-antenna UE as one device in both. Two
mirrored scenes are kept because array configuration in Sionna is a
property of the scene rather than of the device.

Paths are converted to discrete baseband taps by the standard
band-limited sampling of the multipath response~\cite[Eq.~2.34]{tse2005},
\begin{equation}
  \bar h_{\ell} = \sum_{i=0}^{M-1} a_i^{\mathrm{b}}
    \operatorname{sinc}\!\left(\ell - f_s \tau_i\right),
  \qquad \ell = 0,\dots,70,
  \label{eq:taps}
\end{equation}
over a window of $L=71$ taps per antenna pair. The rate appearing in
Eq.~\eqref{eq:taps} is the OAI front-end sample rate $f_s$, not the
nominal channel bandwidth $W$ of Table~\ref{tab:configs}: Sionna's
\texttt{bandwidth} argument is the bandwidth of the receive filter,
which for a complex-baseband channel sampled at the Nyquist rate is the
sampling frequency, and the emulator passes $f_s$. The emulated tap grid and the
grid on which OAI convolves the taps with its sample stream are therefore
the same, which we confirmed by placing a single path at a known delay and
recovering its tap index. \textbf{Delay normalisation is
disabled}, which is distinct from the energy normalisation of
Section~\ref{sec:limits}: the former fixes where a path sits on the
delay axis, the latter the total power of the tap vector, and only the
first is switched off here. Sionna by default shifts the first arrival of
every antenna pair to $\tau=0$, which is convenient for link-level study
and destroys the observable of interest. With absolute delays preserved,
arrival at the farthest element trails the nearest by up to 109\,ns,
the 32.7\,m spread in path length. This is the single most consequential
formatting decision in the emulator.

Because the window holds absolute rather than excess delay it must
accommodate the absolute delay of the farthest element. One tap spans
$c/f_s$ in path length, 4.88\,m and 2.44\,m for the two configurations
(Table~\ref{tab:configs}), so the window reaches $71\,c/f_s$, or 346\,m
and 173\,m. The farthest element lies 56.9\,m away, leaving 290\,m and
116\,m of excess path: ample margin at the narrower bandwidth over a
scene some 190\,m across, less at the wider one.

\subsection{Integration with OAI}

The emulator is a Python process holding the scene and exposing a REST
interface for scenario control. Taps are serialised with FlatBuffers and
published to \texttt{vrtsim} over one nanomsg publish--subscribe socket
per radio device; each message carries the destination device identifier,
the antenna dimensions and the complex tap array, indexed per gNB antenna
so that every antenna port observes its own geometry-consistent channel
with its own absolute delay. The unmodified OAI gNB and nrUE softmodems
run above \texttt{vrtsim} exactly as they would on radio
hardware~\cite{kaltenberger2025oai}. A UE position update
posted to the REST interface triggers a fresh solve on both scenes;
because trajectories are scripted, the ground truth of every channel
realisation is known by construction.
\begin{figure}[t]
  \centering
  \begin{tikzpicture}[
    font=\scriptsize,
    node distance=3.6mm and 3mm,
    box/.style    ={draw, rounded corners=2pt, align=center,
                    minimum height=5.6mm, inner sep=2.6pt, text width=33mm},
    wide/.style   ={box, text width=71.5mm},
    data/.style   ={box, fill=blue!6},
    proc/.style   ={box, fill=black!4},
    wproc/.style  ={wide, fill=black!4},
    oai/.style    ={wide, fill=green!8},
    oaih/.style   ={box, fill=green!8},
    arr/.style    ={-{Stealth[length=1.7mm]}, semithick},
    biarr/.style  ={{Stealth[length=1.7mm]}-{Stealth[length=1.7mm]},
                    semithick},
    grp/.style    ={rounded corners=3pt, draw=black!25, dashed, line width=0.4pt},
    tag/.style    ={font=\scriptsize\itshape, inner sep=1pt, text=black!65}
  ]
    \node[data] (yaml) {YAML scenario\\[-1pt]
      \tiny antenna positions, UE position, $f_c$, $W$};
    \node[data, right=of yaml] (xml) {Mitsuba scene\\[-1pt]
      \tiny \texttt{EURECOM.xml}, ITU materials};
 
    \node[proc, below=of yaml] (dl) {\textbf{DL scene}\\[-1pt]
      \tiny 8 Tx (RU) $\rightarrow$ 1 Rx (UE)};
    \node[proc, below=of xml]  (ul) {\textbf{UL scene}\\[-1pt]
      \tiny 1 Tx (UE) $\rightarrow$ 8 Rx (RU)};
 
    \coordinate (mid) at ($(dl.south)!0.5!(ul.south)$);
    \node[wproc, below=5.2mm of mid] (taps) {Sionna~RT
      \texttt{PathSolver} $\rightarrow$ \texttt{Paths.taps()}\\[-1pt]
      \tiny SBR $+$ image method, \texttt{max\_depth}~10\\[-2pt]
      \tiny 71 absolute-delay taps per antenna pair, unnormalised};
 
    \node[wproc, below=of taps] (tx) {FlatBuffers over nanomsg\\[-1pt]
      \tiny one publish socket per radio device, sliced $[8,1,71]$ /
      $[1,8,71]$};
 
    \coordinate (bl) at ($(tx.south west)+(-1.5mm,-2.6mm)$);
    \coordinate (br) at ($(tx.south east)+( 1.5mm,-2.6mm)$);
    \draw[dashed, black!55, line width=0.5pt] (bl) -- (br);
    \node[tag, anchor=west, fill=white, inner xsep=2pt]
      at ($(bl)+(2mm,0)$) {radio-device boundary};
 
    \node[oai, below=4.6mm of tx] (vrt) {\texttt{vrtsim} virtual radio devices\\[-1pt]
      \tiny 8 RU devices $+$ 1 UE device};
    \node[oaih, below=of vrt.south west, anchor=north west] (gnb)
      {gNB softmodem\\[-1pt] \tiny PHY / MAC / RRC};
    \node[oaih, below=of vrt.south east, anchor=north east] (ue)
      {nrUE softmodem};
    \node[oai, below=of $(gnb.south)!0.5!(ue.south)$, anchor=north] (lmf)
      {5G core $+$ LMF\\[-1pt]
       \tiny NRPPa SRS measurements $\rightarrow$ \mbox{UL-TDoA} solve\\[-2pt]
       \tiny $\rightarrow$ position estimate};
 
    \begin{scope}[on background layer]
      \node[grp, fit=(yaml)(xml)(taps)(tx), inner sep=2.2mm] (gA) {};
      \node[grp, fit=(vrt)(gnb)(ue)(lmf),  inner sep=2.2mm] (gB) {};
    \end{scope}
    \node[tag, anchor=west, fill=white, inner xsep=2pt]
      at ($(gA.north west)+(3mm,0)$) {digital twin (this work)};
    \node[tag, anchor=west, fill=white, inner xsep=2pt]
      at ($(gB.south west)+(3mm,0)$) {unmodified OpenAirInterface};
 
    \coordinate (bus)  at ($(yaml.south)!0.5!(dl.north)$);
    \coordinate (busL) at (yaml.south |- bus);
    \coordinate (busR) at (xml.south  |- bus);
    \draw[semithick] (yaml.south) -- (busL);
    \draw[semithick] (xml.south)  -- (busR);
    \draw[semithick] (busL) -- (busR);
    \draw[arr] (busL) -- (dl.north);
    \draw[arr] (busR) -- (ul.north);
    \draw[arr] (dl.south) -- ++(0,-2.0mm) -| (taps.north);
    \draw[arr] (ul.south) -- ++(0,-2.0mm) -| (taps.north);
    \draw[arr] (taps) -- (tx);
    \draw[arr] (tx.south) -- (vrt.north);
    \draw[biarr] (vrt.south -| gnb.north) -- (gnb.north);
    \draw[biarr] (vrt.south -| ue.north)  -- (ue.north);
    \draw[biarr] (gnb.east) -- (ue.west);
    \draw[biarr] (gnb.south) -- ++(0,-2.4mm) -| (lmf.north);
  \end{tikzpicture}
  \caption{The digital twin replaces the over-the-air RF link at the
    radio-device boundary. Taps cross that boundary in one direction;
    samples cross the vrtsim interface in both, and the nrUE's protocol
    peer is the gNB over the emulated Uu interface. Everything beyond the
    boundary (PHY, MAC, RRC, core and LMF) is unmodified OAI.}
  \label{fig:arch}
\end{figure}
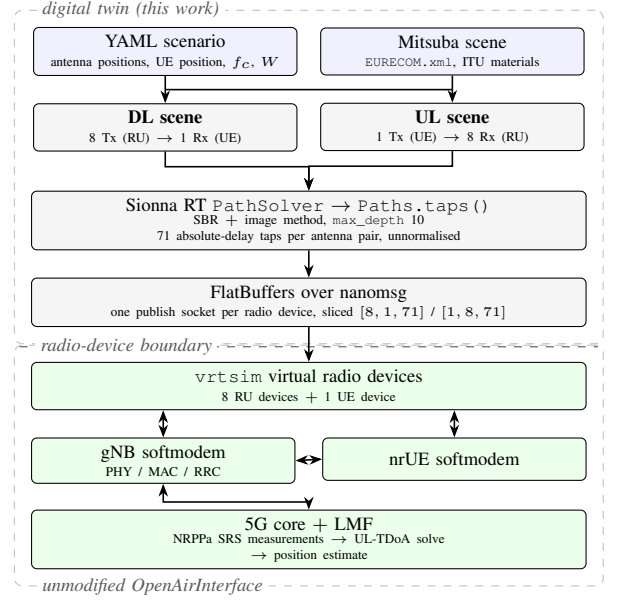

\section{UL-TDoA Positioning and Its Conditioning}
\label{sec:tdoa}

\subsection{Measurement chain}

UL-TDoA locates a UE from differences between the times at which one
transmission reaches receivers of known position~\cite{3gpp38305}.
With the UE at $\mathbf{p}$ transmitting at unknown instant $t_0$ and
TRP~$i$ at $\mathbf{p}_i$ observing arrival at $\hat\tau_i$, each
measurement satisfies $c(\hat\tau_i - t_0) = r_i$, where
$r_i = \lVert \mathbf{p} - \mathbf{p}_i \rVert$ is the range from the UE
to that TRP and $c$ the speed of light. Differencing against a reference
TRP, indexed~0, eliminates the unknown transmit instant $t_0$,
\begin{equation}
  d_i = c\,(\hat\tau_i - \hat\tau_0) = r_i - r_0,
  \qquad i = 1 \ldots N-1 .
  \label{eq:tdoa}
\end{equation}
Eight TRPs give seven constraints on three unknowns, solved by least
squares. The gNB estimates $\hat\tau_i$ per antenna port from the
received SRS and reports it as an uplink relative time of arrival; the
LMF collects the reports over NRPPa~\cite{3gpp38455} with the TRP
coordinates and returns a position through the location-services
interface~\cite{3gpp23273}.

One property of the twin must be stated plainly: the eight emulated
antennas are branches of one gNB sharing a sample clock, with no
per-branch cable delay or phase offset, so the twin is perfectly
synchronised by construction where the real testbed is
not~\cite{ahadi2025experimental}. It does not reproduce the dominant
error source of a physical TDoA deployment.

\subsection{Geometric conditioning}
\label{sec:dop}

A TDoA system does not locate with uniform quality over its coverage
area. The same timing error produces a small position error where the
geometry is favourable and a large one where it is not, and the factor
relating the two follows from the antenna coordinates alone. The
standard treatment is due to Torrieri~\cite{torrieri1984}; we restate it
here because the distinction drawn at the end of this section carries
much of the argument that follows.

Let $\mathbf{u}_i$ denote the unit vector pointing from the UE towards
TRP~$i$. Displacing the UE by a small $\delta\mathbf{p}$ changes the
range $r_i$ by $-\mathbf{u}_i^{\mathsf{T}}\delta\mathbf{p}$, so it
changes the measured range difference $d_i = r_i - r_0$ of
Eq.~\eqref{eq:tdoa} by
$(\mathbf{u}_0 - \mathbf{u}_i)^{\mathsf{T}}\delta\mathbf{p}$. Stacking
the $N-1$ differences gives $\delta\mathbf{d} = \mathbf{A}\,
\delta\mathbf{p}$, where the $(N-1)\times 3$ matrix
\begin{equation}
  \mathbf{A} = \big[\,\mathbf{u}_1 - \mathbf{u}_0,\;\ldots,\;
                     \mathbf{u}_{N-1} - \mathbf{u}_0\,\big]^{\mathsf{T}}
  \label{eq:jacobian}
\end{equation}
depends only on where the TRPs sit relative to the UE. Solving in the
least-squares sense inverts this relation, and treating the
range-difference errors as uncorrelated with common variance $\sigma^2$
gives a position covariance of $\sigma^2\mathbf{Q}$ with
$\mathbf{Q} = (\mathbf{A}^{\mathsf{T}}\mathbf{A})^{-1}$.

This is the conventional formulation~\cite{torrieri1984}, and it is also
the estimator OAI implements, which performs an unweighted least-squares
solve. The dilution figures are used below to compare conditioning
between positions and between antenna layouts rather than as an exact
error-propagation coefficient.

Every dilution figure is a square root of a sum of diagonal entries of
$\mathbf{Q}$ and is therefore dimensionless: the number by which one
metre of range-difference error becomes metres of position error at that
point. The OAI solver estimates the height together with the horizontal
coordinates, so both figures we use come from the same three-dimensional
solve:
\begin{equation}
  \mathrm{HDOP} = \sqrt{Q_{xx}+Q_{yy}}, \qquad
  \mathrm{VDOP} = \sqrt{Q_{zz}}.
  \label{eq:dop}
\end{equation}
HDOP sums only the horizontal variances, since it describes horizontal
accuracy, and the vertical variance of the same solve is reported
separately as VDOP. The height column nonetheless enters HDOP through the
inversion of the full $3\times 3$ Gram matrix, so a poorly observed
height inflates HDOP even when the horizontal geometry is good, which is
the effect Section~\ref{sec:results} turns on.

\section{Experimental Setup}
\label{sec:setup}

\subsection{Two scenes}

Because TDoA error depends on both array geometry and propagation
environment, an error observed in one scene cannot be attributed to
either. We therefore instantiate the same array in two scenes: EURECOM,
reproducing the deployment site, and the Munich scene distributed with
Sionna~RT, an unrelated dense-urban model. The Munich coordinates are the
image of the EURECOM coordinates under a distance-preserving transform,
so every quantity the solver's conditioning depends on is carried across
unchanged: element spacings, collinearity, orthogonality of the two axes
and the 42.19\,m centroid separation. The array geometry is therefore held fixed
while the propagation environment changes. The test positions are not
the image of one another under the same transform: each set was chosen
independently within its scene, so the two sweeps differ in position as
well as in environment and are compared in aggregate rather than as a
controlled substitution of one variable.

Sixteen test positions per scene form two tracks running away from the
array at $z=1.7$\,m, spanning $\mathrm{HDOP}$ from 2.24 to 7.79
(EURECOM) and 2.06 to 7.36 (Munich). The geometry therefore predicts,
before any measurement, that error at the far end of each track should
be roughly three and a half times that at the near end.

\subsection{Two bandwidths}

\begin{table}[t]
  \centering
  \caption{The two configurations. Both place the correlation peak on
    the same 0.61\,m bin spacing; oversampling changes the spacing at
    which a peak may be reported, not the physical delay resolution,
    which follows the occupied bandwidth.}
  \label{tab:configs}
  \footnotesize
  \begin{tabular}{@{}lrr@{}}
    \toprule
     & Config.~A & Config.~B \\
    \midrule
    Resource blocks (30\,kHz SCS)  & 106        & 273 \\
    Channel bandwidth $W$          & 40\,MHz    & 100\,MHz \\
    Occupied bandwidth             & 38.16\,MHz & 98.28\,MHz \\
    OAI sample rate $f_s$          & 61.44\,MHz & 122.88\,MHz \\
    Path length per tap            & 4.88\,m    & 2.44\,m \\
    \addlinespace
    SRS IDFT oversampling factor   & 8          & 4 \\
    Delay-estimate bin             & 2.03\,ns   & 2.03\,ns \\
    Path length per bin            & 0.61\,m    & 0.61\,m \\
    \midrule
    Carrier / band                 & \multicolumn{2}{c}{3.6192\,GHz, n78} \\
    Antenna ports                  & \multicolumn{2}{c}{8 Tx, 8 Rx} \\
    Tap window                     & \multicolumn{2}{c}{71 taps} \\
    Positions per scene            & \multicolumn{2}{c}{16, one pass each} \\
    \bottomrule
  \end{tabular}
\end{table}

Table~\ref{tab:configs} gives the two configurations. It should be read
as a comparison of two complete operating points rather than as an
isolation of propagation bandwidth: changing the PRB count changes the
occupied bandwidth, the sample rate, the tap spacing and therefore the
physical duration of the 71-tap window, all at once. We did not hold the
absolute-delay window constant, so a residual contribution from the
shorter window of configuration~B cannot be excluded.

Two quantisations are worth separating out. The tap grid of the emulated
channel places a floor on where an arrival can be positioned: at 4.88\,m
per tap in configuration~A it is coarser than the bin on which the peak
is reported, and although it is finer than the occupied bandwidth can
resolve on average, Section~\ref{sec:results} shows one poorly
conditioned position at which it appears to bind. The receiver's
delay-estimate bin is independent of propagation bandwidth and, at the
released oversampling factor, comparable to the tap spacing of
configuration~B. Raising the factor to 8 and 4 respectively brings both
to 2.03\,ns (0.61\,m). This equalises the spacing at which a correlation
peak may be reported; it does not equalise the physical delay
resolution, which follows the occupied bandwidth and necessarily differs
between the two. The asymmetry is
forced: the arrays the constant dimensions are stack-allocated in the L1
receive thread, and factor~8 at 273 PRB exhausts the stack.

The LMF used throughout writes its linearised TDoA system in three
dimensions, carries the range to the reference TRP as an explicit
unknown and solves by QR least squares; the gNB arrival-time estimator
reads the peak of its inverse transform on the oversampled grid. Both are
available on branches of the respective repositories.

Six further parameters differ between the two gNB configuration files.
Four govern random access only. The two uplink power-control targets
differ by 6\,dB, so we tested them directly: re-running configuration~A
over the full EURECOM track at the configuration~B value reproduced all
sixteen fixes exactly, as expected of a noiseless correlation whose peak
does not move with transmit power.

\subsection{Determinism and what the statistics mean}

The emulator is deterministic and its noise power is zero. A Munich sweep
repeated immediately after itself reproduced all sixteen fixes to the last
recorded digit. One pass per configuration is therefore sufficient, but it
follows that the errors reported here are biases rather than noise
realisations: at a given position the pipeline returns the same wrong
answer every time, and averaging cannot reduce it. All quantiles below are
taken over the sixteen positions of a track, not over independent trials.
They describe how error varies with geometry, not how it would vary under
repeated measurement at one point.

\section{Results}
\label{sec:results}

\subsection{Horizontal accuracy}

\begin{table}[t]
  \centering
  \caption{Aggregate error over sixteen positions per sweep, two-height
    array. $e_{\mathrm{2D}}$ is horizontal; $\Delta z$ is the signed
    height error.}
  \label{tab:summary}
  \footnotesize
  \begin{tabular}{@{}llrrrrrr@{}}
    \toprule
    & & \multicolumn{5}{c}{$e_{\mathrm{2D}}$ (m)} & $\Delta z$ \\
    \cmidrule(lr){3-7}\cmidrule(lr){8-8}
    Scene & $W$ & Mean & Med. & 90th & Max & $<\!1.5$ & RMS \\
    \midrule
    EURECOM & 40  & 1.28 & 1.20 & 1.77 & 3.81  & 11/16 & 3.50 \\
    EURECOM & 100 & 1.15 & 0.95 & 1.80 & 2.73  & 12/16 & 3.70 \\
    Munich  & 40  & 2.62 & 0.98 & 5.39 & 19.07 & 11/16 & 2.31 \\
    Munich  & 100 & 1.04 & 0.71 & 2.32 & 3.70  & 13/16 & 2.19 \\
    \bottomrule
  \end{tabular}
\end{table}

Table~\ref{tab:summary} gives the aggregate result with the array as
deployed. Median error is 1.20\,m and 0.95\,m at EURECOM and 0.98\,m and
0.71\,m in Munich, and eleven to thirteen of sixteen positions fall
inside the 1.5\,m target. One fix stands apart, 19.07\,m at Munich position~1 at 40\,MHz. That
position is the worst conditioned on its track
($\mathrm{HDOP}=6.96$ against a median of 3.28) and the implied
range-difference error of 2.74\,m is roughly half the 4.88\,m tap
spacing, so the first arrival has most likely been assigned to a
neighbouring tap and poor conditioning has amplified the result; the same
position returns 2.61\,m at 100\,MHz. Excluding it, no fix exceeds
7\,m and most are under 2\,m. Means exceed medians in every sweep, which
is what a geometry-driven error distribution looks like.

\begin{figure}[t]
  \centering
  \includegraphics[width=\columnwidth]{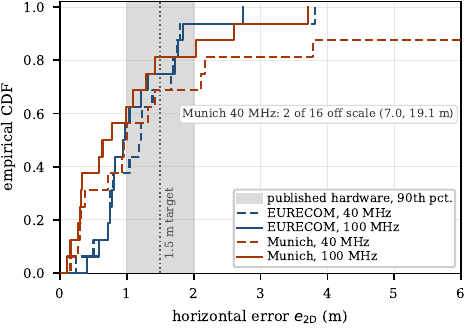}
  \caption{Horizontal error distribution with the two-height array,
    sixteen positions per sweep. The shaded band is the 1--2\,m 90th
    percentile reported for the physical
    deployment~\cite{ahadi2025experimental}. Widening the carrier shifts
    both scenes left, and the gain is largest in the tail rather than at
    the median.}
  \label{fig:cdf}
\end{figure}

\subsection{The error follows the geometry}

Under the model of Section~\ref{sec:dop} the error at each position should
be a range-difference error $\sigma$ amplified by the local dilution.
Inverting this, by taking the median of
$e_{\mathrm{2D}}/\mathrm{HDOP}$ over each sweep, estimates $\sigma$
from the measurements themselves: 0.307, 0.306, 0.308 and 0.238\,m for
EURECOM at 40 and 100\,MHz and Munich at 40 and 100\,MHz respectively.

Three of the four agree to within 1\%, across two entirely different
propagation environments and two bandwidths. The common value of
0.31\,m corresponds to 1.02\,ns, which is one step of the UL-RTOA
reporting granularity $T_c/2 = 1.017$\,ns. The residual measurement error
is therefore roughly one least significant bit of the format in which the
gNB reports it, and the pipeline sits close to the floor that
representation permits; the lower Munich value is consistent with that
floor being approached from above.

Partitioning positions at the median $\mathrm{HDOP}$ of each track
shows the amplification directly. In Munich the correlation between
$e_{\mathrm{2D}}$ and $\mathrm{HDOP}$ is 0.71 and 0.66 at the two
bandwidths, and the mean error in the worse-conditioned half exceeds the
better-conditioned half by $5.8\times$ at 40\,MHz. At EURECOM the
correlation is weak (0.44 and 0.00) for an instructive reason: errors
there span only 0.24 to 3.81\,m, so nearly every position sits close to
the measurement floor and there is little for the geometry to explain.

\subsection{Effect of channel bandwidth}

Widening the carrier from 40 to 100\,MHz reduces the median horizontal
error from 1.20 to 0.95\,m at EURECOM and from 0.98 to 0.71\,m in Munich.
Those aggregate figures conceal what actually changed. The same sixteen
positions are swept in both configurations, so each fix can be paired
with its counterpart, and doing so gives five improvements, seven
degradations and four unchanged positions at EURECOM, and seven, six and
three in Munich. The median paired change is zero in both scenes to two
decimal places. The wider carrier does not make a typical position more
accurate.

What it does is truncate the upper tail. The largest error falls from
3.81 to 2.73\,m at EURECOM and from 19.07 to 3.70\,m in Munich, and the
Munich 90th percentile falls from 5.39 to 2.32\,m. The aggregate medians
improve because that truncation reshapes the distribution, not because
positions individually improve. The mean makes the same point more
starkly: the Munich mean falls from 2.62 to 1.04\,m, but with the two
fixes above 6\,m excluded it falls only from 1.13 to 0.73\,m.

This is predicted rather than disappointing. If the pipeline already sits
at the floor set by the reporting granularity, most positions have
nothing for additional bandwidth to recover, and only positions whose
delay estimate was genuinely mis-resolved can benefit. Splitting each
track at its median $\mathrm{HDOP}$ shows exactly that: in the
worse-conditioned half of the Munich track the mean falls from 4.47 to
1.44\,m and five of eight positions improve, while in the
better-conditioned half it falls only from 0.77 to 0.63\,m and two of
eight improve. At EURECOM, where the whole track is comparatively well
conditioned, the better-conditioned half is marginally \emph{worse} at
100\,MHz, 1.14 against 1.09\,m.

The supported conclusion is therefore narrower than a general bandwidth
effect and is specific to these two configurations. Additional
propagation bandwidth does not lower the error at positions already
limited by the reporting format; it reduces the frequency and severity of
the large errors that arise where the geometry amplifies a mis-resolved
delay. A coverage area that is uniformly well conditioned would gain
little from the wider carrier, while one with poorly conditioned regions
would gain in the tail rather than at the median.

The bandwidth has \emph{no} useful effect on the vertical component: RMS
height error moves from 3.50 to 3.70\,m at EURECOM and 2.31 to 2.19\,m
in Munich, in opposite directions and by under 6\%, against a
$2.5\times$ change in occupied bandwidth. These are deterministic
outcomes at sixteen fixed positions rather than noise realisations, so
the residual movement is not sampling error; what matters is that it is
small and inconsistent in sign.

\subsection{Why the array cannot see height}

A quantity that does not respond to a $2.5\times$ increase in measurement
bandwidth is unlikely to be limited by the measurement. The geometry
offers an explanation, which this section tests first against the array
coordinates and then by changing them.

All eight antennas occupy exactly two heights. The cost is clearest in
the system the solver actually forms. Squaring $r_i = r_0 + d_i$ and
subtracting the reference equation removes the quadratic term in
$\mathbf{p}$ and leaves a system linear in $(x,y,z,r_0)$, with row~$i$
reading $2(\mathbf{p}_i - \mathbf{p}_0)^{\mathsf{T}}\mathbf{p} +
2 d_i r_0 = \lVert\mathbf{p}_i\rVert^2 -
\lVert\mathbf{p}_0\rVert^2 - d_i^2$. This is what the LMF implements,
and it is a different object from the unit-vector Jacobian of
Eq.~\eqref{eq:jacobian}: that one linearises about a position estimate
and has dimensionless rows, this one is exact and carries rows in
metres. Section~\ref{sec:dop} uses the first because dilution is defined
for it; the argument here uses the second because it exposes the height
dependence directly. Its height column, in metres and with TRP~1 as
reference, is
\begin{equation}
  \big[\,0,\; 0,\; 0,\; 21.6,\; 21.6,\; 21.6,\; 21.6\,\big]^{\mathsf{T}}.
  \label{eq:zcol}
\end{equation}
Three of the seven equations carry no height information at all, because
those TRPs sit at exactly the reference height; the remaining four repeat
one coefficient. Height therefore enters through a single linear
combination of the measurements, the mean offset between the street
group and the elevated group, and that combination is further confounded
with $r_0$, which the solver estimates simultaneously. The four elevated
antennas average down the variance of that one combination but supply no
second direction along which $z$ can be resolved. The two groups are
separated by 10.8\,m vertically while the UE stands 20 to 60\,m away, so
the array judges height from a vertical baseline two to six times shorter
than its working range.

The second consequence is more expensive: the cost is not confined to
the vertical coordinate. Median $\mathrm{HDOP}$ is 3.53 at EURECOM
and 3.28 in Munich, and Table~\ref{tab:dop} shows it halving once
height becomes observable, with the horizontal coordinates untouched.
That factor of two is the price of estimating a coordinate the array can
barely observe: information that would otherwise constrain $x$ and $y$ is
spent on $z$.

The model can be tested without a free parameter. Taking the
$\hat\sigma$ of each sweep, estimated above from its \emph{horizontal}
errors alone, the normalised height residual
$\tilde z_k = \Delta z_k / (\mathrm{VDOP}_k\,\hat\sigma)$ should have zero
mean and unit standard deviation if the height error is nothing but the
same measurement error amplified by vertical geometry. Over the sixteen
EURECOM positions at 40\,MHz its mean is $-0.07$ and its standard
deviation 1.02.

\subsection{Answering a question the testbed cannot}

The height failure is a property of the layout, so we changed the layout.
Spreading heights evenly along each sub-array is a poor choice, because
elements within a group are only 5 to 9\,m apart horizontally and a
gradual progression leaves the baselines nearly horizontal; what tilts a
short baseline is a large height difference between \emph{adjacent}
elements. Alternating heights within each group (Table~\ref{tab:dop},
caption) inclines every baseline by more than $50^{\circ}$. The height
column of Eq.~\eqref{eq:zcol} becomes
$[\,20.6,\,0,\,20.6,\,21.6,\,60.6,\,21.6,\,60.6\,]^{\mathsf{T}}$: six
non-zero coefficients taking three distinct values that span a factor of
three, rather than a single value repeated four times.

\begin{table}[t]
  \centering
  \caption{Median dilution over the sixteen test positions, before and
    after the height change. Heights are
    $(1.7,1.7,1.7,1.7,12.5,12.5,12.5,12.5)$\,m before and
    $(1.7,12.0,1.7,12.0,12.5,32.0,12.5,32.0)$\,m after; horizontal
    coordinates are untouched. Predicted $\Delta z$ is
    $\hat\sigma\,\mathrm{VDOP}$ at $\hat\sigma = 0.31$\,m.}
  \label{tab:dop}
  \footnotesize
  \begin{tabular}{@{}llrrr@{}}
    \toprule
    Scene & Layout & HDOP & VDOP & Pred.\ $\Delta z$ \\
    \midrule
    EURECOM & two-height     & 3.53 & 10.65 & 3.30\,m \\
    EURECOM & height-diverse & 1.77 &  2.12 & 0.66\,m \\
    Munich  & two-height     & 3.28 & 10.26 & 3.18\,m \\
    Munich  & height-diverse & 1.63 &  2.04 & 0.63\,m \\
    \bottomrule
  \end{tabular}
\end{table}

Two features of Table~\ref{tab:dop} matter. VDOP falls by a factor of
five at both sites, and $\mathrm{HDOP}$ by a factor of two, although
the horizontal antenna coordinates are identical in both layouts: the
horizontal gain comes purely from the solver no longer spending
information on an unobservable coordinate. In the twin this costs one
configuration file; on the physical deployment it would require
remounting four antennas.

\begin{table}[t]
  \centering
  \caption{Aggregate error before and after the height change.}
  \label{tab:diverse}
  \footnotesize
  \begin{tabular}{@{}llrrrrr@{}}
    \toprule
    & & \multicolumn{4}{c}{$e_{\mathrm{2D}}$ (m)} & $\Delta z$ \\
    \cmidrule(lr){3-6}\cmidrule(lr){7-7}
    Scene / $W$ & Layout & Mean & Med. & 90th & Max & RMS \\
    \midrule
    EUR.\ 40  & two-height     & 1.28 & 1.20 & 1.77 &  3.81 & 3.50 \\
              & height-div.    & 1.24 & 1.03 & 2.62 &  3.52 & \textbf{0.85} \\
    \addlinespace
    EUR.\ 100 & two-height     & 1.15 & 0.95 & 1.80 &  2.73 & 3.70 \\
              & height-div.    & 1.54 & 0.85 & 4.47 &  5.73 & \textbf{1.02} \\
    \addlinespace
    Mun.\ 40  & two-height     & 2.62 & 0.98 & 5.39 & 19.07 & 2.31 \\
              & height-div.    & 1.22 & 0.84 & 2.34 &  3.79 & \textbf{0.95} \\
    \addlinespace
    Mun.\ 100 & two-height     & 1.04 & 0.71 & 2.32 &  3.70 & 2.19 \\
              & height-div.    & 1.29 & 0.56 & 3.52 &  6.20 & \textbf{0.63} \\
    \bottomrule
  \end{tabular}
\end{table}

Both scenes were re-swept over the same sixteen positions with the new
layout and everything else unchanged (Table~\ref{tab:diverse}). The
vertical improvement is large and consistent: RMS height error falls from
3.50 to 0.85\,m at EURECOM and 2.31 to 0.95\,m in Munich at 40\,MHz, with
comparable factors at 100\,MHz, and the worst single height error over all
four sweeps falls from 7.44 to 2.05\,m. Height is now recovered to roughly
the accuracy of the horizontal coordinates, which
Table~\ref{tab:dop} predicts and which the two-height array cannot achieve
at any bandwidth. The bandwidth ordering survives the layout change, so it
is not an artefact of the original geometry.

\begin{figure}[t]
  \centering
  \includegraphics[width=\columnwidth]{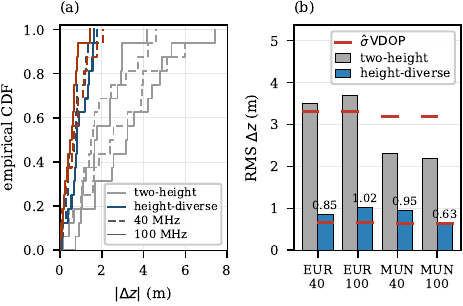}
  \caption{(a) Distribution of the height error over sixteen positions
    per sweep, before (grey) and after (colour) the height change; dashed
    is 40\,MHz, solid 100\,MHz. The two-height array spreads to 7\,m,
    the height-diverse array stays inside 2\,m. (b) RMS height error at
    the four operating points against the $\hat\sigma\,\mathrm{VDOP}$
    prediction of Table~\ref{tab:dop}. Horizontal coordinates are
    identical in both layouts.}
  \label{fig:height}
\end{figure}

The horizontal result deserves to be stated plainly rather than folded
into the vertical success. The median improves at all four operating
points (1.20$\to$1.03, 0.95$\to$0.85, 0.98$\to$0.84, 0.71$\to$0.56\,m),
but the mean and 90th percentile do not: at EURECOM 100\,MHz the 90th
percentile rises from 1.80 to 4.47\,m. Table~\ref{tab:dop} predicted a
factor-of-two horizontal gain and it was not realised. Re-estimating
$\hat\sigma$ on the new sweeps explains why: the implied range-difference
error rises from 0.307 to 0.575\,m at EURECOM and 0.308 to 0.535\,m in
Munich. The geometry improved by the predicted factor and the measurement
degraded by approximately the same factor; the two cancel in the
horizontal and leave a large net gain in the vertical, where the geometric
factor is five rather than two. The most likely cause is that raising four
elements to 32\,m places them above the surrounding rooftops in both
scenes, lengthening their paths and putting some building faces at grazing
incidence. We flag this as a hypothesis: confirming it requires the
per-TRP delay profiles, which we did not record.

\subsection{Comparison with the physical testbed}

The same pipeline on the physical GEO-5G testbed reaches 1--2\,m in
90\% of cases~\cite{ahadi2025experimental}. The defensible comparison is
the same statistic at the closest configuration: the EURECOM scene at
100\,MHz, which reproduces the deployment site, gives a 90th percentile
of 1.80\,m, inside the published range. The Munich 90th percentile is
2.32\,m and the 40\,MHz sweeps reach 5.39\,m, so the agreement is
specific to the matched case rather than general, and our tracks
deliberately extend into the poorly conditioned region beyond the end of
the array, where the published evaluation did not go. Medians appear in
Table~\ref{tab:summary} but are not comparable to a 90th-percentile
figure and should not be read against them.

Differences run in both directions. In the twin's favour: perfect
synchronisation, a sparser impulse response and no receiver noise beyond
OAI's own. Against it: the published results apply dedicated ToA/TDoA
filtering across repeated measurements whereas each figure here is a
single fix, and the emulated tap grid places a floor under the arrival
estimate that a physical receiver does not share in the same form. The
comparison is indicative rather than like-for-like; within that limit the
twin reproduces the accuracy regime of the system it models.

\section{Limitations}
\label{sec:limits}

\emph{Statistical weight.} The emulator is deterministic with zero noise
power, so repeated sweeps confirm repeatability without sampling a random
process. Sixteen positions on two tracks is a small, non-random sample,
and the 90th percentile is estimated from effectively two points. Genuine
error statistics require a noise realisation or a perturbation of the UE
position within a neighbourhood; neither was done here.

\emph{Bandwidth as a variable.} The two configurations differ in
occupied bandwidth, sample rate, tap spacing and the physical duration
of the tap window together. Attributing the observed improvement to
propagation bandwidth alone would require holding the absolute-delay
window constant and recording per-TRP delay profiles, neither of which
we did.

\emph{Model and scope.} Diffuse scattering, mutual coupling, per-branch RF
impairments and antenna pattern detail are absent, and energy
normalisation discards absolute path loss. The emulator produces
independent static snapshots, so nothing here speaks to mobility, Doppler
or tracking. The contribution of the distributed geometry relative to a
co-located one is assumed rather than measured, and the height-diverse
layout was evaluated at a single assignment, so the trade-off between
vertical observability and link quality is signalled but not mapped.

\section{Conclusion}

We built a ray-traced digital twin of a distributed eight-antenna 5G
testbed and used it to drive an unmodified OAI UL-TDoA positioning
pipeline. Modelling each antenna as an independent radio device preserves
the per-antenna absolute delays a TDoA estimator consumes, and with them
preserved the pipeline recovers position to a median horizontal error of
0.71--1.20\,m.

The twin then answered a question the hardware cannot. Vertical accuracy
proved insensitive to bandwidth because it is limited by an array that
occupies only two heights, a limit computable from the antenna coordinates
before any measurement is taken; a height-diverse layout cut RMS height
error by a factor of four at the cost of one configuration file. A
reference carrying no error of its own lets a deployment be interrogated
in ways a surveyed site cannot easily support.

The per-position output of every sweep reported here, together with the
gNB configuration used to produce it, is available
at~\cite{dt_results_repo}, so that every figure and table above can be
recomputed from the raw fixes.

Future work follows directly from Section~\ref{sec:limits}: adding a noise
realisation and perturbation sweeps to obtain genuine error statistics,
recording per-TRP arrival times to decompose error into timing and
geometric components, mapping the height/link-quality trade-off over a
range of layouts, and comparing the emulated channel against measurements
from the physical site.

\bibliographystyle{IEEEtran}
\bibliography{wons2027_refs}

\end{document}